# Dimensional crossover in a layered ferromagnet detected by spin correlation driven distortions

A. Ron[1,2], E. Zoghlin[3], L. Balents[4], S. D. Wilson[3], & D. Hsieh[1,2*]

[1]*Department of Physics, California Institute of Technology, Pasadena, California 91125, USA.*

[2]*Institute for Quantum Information and Matter, California Institute of Technology, Pasadena, California 91125, USA.*

[3]*Materials Department, University of California, Santa Barbara, California 93106, USA.*

[4]*Kavli Institute for Theoretical Physics, University of California, Santa Barbara, California 93106, USA.*

*dhsieh@caltech.edu

**Magneto-elastic distortions are commonly detected across magnetic long-range ordering (LRO) transitions. In principle, they are also induced by the magnetic short-range ordering (SRO) that precedes a LRO transition, which contains information about short-range correlations and energetics that are essential for understanding how LRO is established. However these distortions are difficult to resolve because the associated atomic displacements are exceedingly small and do not break symmetry. Here we demonstrate high-multipole nonlinear optical polarimetry as a sensitive and mode selective probe of SRO induced distortions using CrSiTe$_3$ as a testbed. This compound is composed of weakly bonded sheets of nearly isotropic ferromagnetically interacting spins that, in the Heisenberg limit, would individually be impeded from LRO by the Mermin-Wagner theorem. Our results show that CrSiTe$_3$ evades this law via a two-step crossover from two- to three-dimensional magnetic SRO, manifested through two successive and previously undetected totally symmetric distortions above its Curie temperature.**

Ferromagnetic (FM) semiconductors belonging to the transition metal trichalcogenide family have recently been shown to be promising starting materials for realizing monolayer ferromagnets by exfoliation[1,2]. However predicting the viability of the ferromagnetic long-range ordered state in the 2D limit relies on first understanding how FM LRO is established in the 3D bulk crystals, which is often unclear. A case in point is CrSiTe$_3$[3–8], which consists of *ABC* stacked sheets of Cr$^{3+}$ (spin-3/2) moments arranged in a honeycomb network (Fig. 1). Each Cr atom is coordinated by six Te atoms that form an almost perfect octahedron[7], giving rise to a near isotropic (Heisenberg) spin state and dominant FM nearest neighbor exchange interactions ($J_{ab} < 0$) owing to the near 90° Cr-Te-Cr bond angle. This is corroborated by inelastic neutron scattering experiments[7] on bulk CrSiTe$_3$, which report a relatively feeble easy-axis (Ising) anisotropy strength ($D/J_{ab} < 2\,\%$). According to the Mermin-Wagner theorem[9], LRO should be forbidden in a strictly 2D Heisenberg system. Therefore the finite value of the Curie temperature ($T_c \sim 31$ K) in bulk CrSiTe$_3$ (Fig. 1) must either be driven by the weak spin anisotropy or by a weak interlayer coupling that mediates a crossover from 2D to 3D character. A dimensional crossover can in principle be uncovered by tracking the spatial anisotropy of short-range spin correlations using magnetic neutron and x-ray scattering techniques. However the requirement of nearly ideal bulk crystals and the difficulty of detecting and integrating diffuse magnetic scattering is restrictive and currently renders these techniques inoperable on exfoliated nanoscale thick sheets. Hence this mechanism is yet to be verified in $CrSiTe_3$ or related compounds[1].

An alternative route to measuring short-range spin correlations is through their effects on the crystal lattice. The magnetic energy of an insulating system is given by the thermal expectation value of its magnetic Hamiltonian $\mathcal{H}_m = J_{ij} \sum_{\langle i,j \rangle} \vec{S}_i \cdot \vec{S}_j$, which contains the short-range spin correlator $\langle \vec{S}_i \cdot \vec{S}_j \rangle$ as well as the exchange interaction $J_{ij}$ between spins at sites *i* and *j*. Upon onset of magnetic SRO, it may be energetically favorable for the system to re-adjust the distances and bonding angles between atoms that mediate $J_{ij}$ in order to lower its magnetic energy, at the expense of some gain in elastic energy. Measuring such magneto-elastic distortions therefore yields information about spin correlations along various directions and, for simple low dimensional Hamiltonians, can even provide quantitative values of the $\langle \vec{S}_i \cdot \vec{S}_j \rangle$ function[10], which is difficult to obtain by neutron scattering because only a limited range of its spatio-temporal Fourier components are accessed. However, SRO induced distortions are extremely hard to resolve because they are minute by virtue of $\langle \vec{S}_i \cdot \vec{S}_j \rangle$ being small, and because they generally do not break any lattice symmetries. A suitable probe must therefore be sensitive to and able to distinguish between different totally symmetric distortions (i.e. different basis functions of the totally symmetric irreducible representation). This suggests that examining the nonlinear high rank tensor responses of a crystal may be a promising approach.

Optical second harmonic generation (SHG), a frequency doubling of light produced by its nonlinear interactions with a material, is governed by high rank (> 2) susceptibility tensors that are sensitive to many degrees of freedom in a crystal. Traditionally SHG has been exploited as a symmetry sensitive probe because the leading order electric dipole susceptibility necessarily vanishes if the system possesses a center of inversion. This makes

SHG particularly powerful for studying surfaces of centrosymmetric crystals[11], and for identifying bulk symmetry breaking phase transitions through the appearance of additional, high symmetry forbidden, tensor elements[12–15]. In principle, SHG can also be utilized to study symmetry preserving distortions[16] by examining their subtle effects on the existing symmetry allowed tensor elements. However this potential capability is highly underexplored, in part due to the technical demand of simultaneously tracking small changes across an entire set of allowed tensor elements.

Recently we managed to surmount this challenge by developing a rotating scattering plane based SHG polarimetry technique[17]. In these experiments, linear (either P or S) polarized light of frequency ω is focused obliquely onto the surface of a bulk single crystal. The intensity of either the P or S component of reflected light at frequency 2ω is then measured as a function of the angle (φ) that the scattering plane is rotated about the $c$-axis (Fig. 2a), which allows a multitude of SHG susceptibility tensor elements to be sampled. By collecting these rotational anisotropy (RA) patterns with different polarization combinations, a complete set of SHG susceptibility tensor elements can typically be uniquely determined. Here we apply this technique to track the magnitudes of all of the symmetry allowed SHG susceptibility tensor elements of CrSiTe$_3$ as a function of temperature. Evidence of previously undetected structural distortions are observed above $T_c$ at $T_{2D}$ ~ 110 K and $T_{3D}$ ~ 60 K. Using a hyper-polarizable bond model, we are able to attribute the distortions at $T_{2D}$ and $T_{3D}$ to displacements along different totally symmetric normal mode coordinates, which are consistent with an onset of intralayer and interlayer spin correlations respectively.

The full temperature evolution of the RA patterns acquired from the (001) surface of CrSiTe$_3$ under select polarization geometries is displayed in Figure 2b. We first note that a finite weak SHG intensity is present at all temperatures despite previous work showing that CrSiTe$_3$ always retains a centrosymmetric structure with $\bar{3}$ point group symmetry[18]. This suggests that the SHG originates from a higher multipole process such as electric quadrupole (EQ) radiation, which is governed by a fourth rank susceptibility tensor $\chi_{ijkl}$ that has only 8 independent non-zero elements ($\chi_{xxxz}$, $\chi_{xxyy}$, $\chi_{xxzz}$, $\chi_{yxxx}$, $\chi_{yyyz}$, $\chi_{zzxx}$, $\chi_{zzxy}$, $\chi_{zzzz}$) after accounting for the symmetries of the $\bar{3}$ point group and the degeneracy of the incident electric fields[19]. Expressions for the RA SHG intensity $I(2\omega) \propto \left|\hat{e}_i^{2\omega}(\varphi)\chi_{ijkl}\hat{e}_j^{\omega}(\varphi)\kappa_k(\varphi)\hat{e}_l^{\omega}(\varphi)\right|^2 I_0^2(\omega)$ derived under these conditions (here $\hat{e}$ are the polarization directions, $\kappa$ is the incident wave vector and $I_0$ is the incident intensity; see Supplementary Note 1) indeed produce excellent fits to our set of RA patterns at any given temperature and allow us to uniquely determine the values of $\chi_{ijkl}$ at each temperature. In contrast, other possible allowed SHG processes such as surface electric dipole or bulk magnetic dipole radiation cannot reproduce the RA data and are thus treated as negligibly small (see Supplementary Note 2).

From the raw RA data (Fig. 2b) we can clearly discern the bulk 3-fold rotational symmetry of CrSiTe$_3$ and, as expected, we observe no change in symmetry as a function of temperature. Yet the absolute and relative intensities of the various features do undergo changes upon cooling, which must encode symmetry preserving distortions. Most notably, there is a dramatic increase of intensity below $T_c$ that, as we will show later on, arises from LRO induced magneto-elastic distortions that have previously been detected by optical absorption[20], Raman scattering[20] and x-ray diffraction[18], and are also captured by our dilatometry measurements (Fig. 1a). Surprisingly however, we find that the RA patterns continue to subtly evolve even far above $T_c$. In SP polarization geometry for example (Fig. 2c), representative RA patterns at 140 K, 80 K and 40 K have qualitatively different shapes, indicating that the magnitude of the $\chi_{ijkl}$ elements change non-uniformly with temperature.

The temperature ($T$) dependence of each of the 8 individual $\chi_{ijkl}$ elements was extracted through the aforementioned fitting procedure (Fig. 2c). From every $\chi_{ijkl}(T)$ curve, we subtracted a high temperature background using the data above 150 K, where the shapes of the RA patterns have ceased evolving (see Supplementary Note 3). Figure 3 shows the complete set of background subtracted curves $\Delta\chi_{ijkl}(T)$ that have all been normalized to their low temperature values. Three distinct sets of behavior are clearly resolved. Below a characteristic temperature $T_{2D}$ ~ 110 K, the *xxxz* and *yyyz* elements alone start to grow in tandem (Fig. 3a). Then below a second characteristic temperature $T_{3D}$ ~ 60 K, a solitary *zzzz* element begins to grow (Fig. 3b). The temperature dependence of the former and latter set of elements are sub-linear above $T_c$ and scale with classical calculations of the nearest neighbor intralayer and interlayer spin correlators respectively (see Supplementary Note 5). By contrast, below $T_c$ the remaining 5 elements turn up with an order parameter like temperature dependence indicative of a phase transition (Fig. 3c), with a critical exponent twice that reported[7] for the magnetization ($2\beta \approx 0.3$). Since magneto-elastic distortions scale like the square of the magnetic order parameter, this further confirms that $\chi_{ijkl}$ is probing the lattice degrees of freedom. This also shows that $\chi_{ijkl}$ is a time-reversal invariant *i*-tensor[19], which naturally explains why our measurements are insensitive to magnetic domains[21].

To understand the microscopic origin of the features in $\Delta\chi_{ijkl}(T)$, we appeal to a simplified hyper-polarizable bond model[22], which treats the crystal as an array of charged anharmonic oscillators centered at the chemical bonds and constrained to only move along the bond directions. The nonlinear polarizability of each oscillator is calculated by solving classical equations of motion, and then appropriately summed together to form the total nonlinear susceptibility. Recently an expression for the EQ SHG susceptibility was derived using this model[23] and was found to take the form $\chi_{ijkl} \propto \sum_n \alpha_\omega \alpha_{2\omega} (\hat{b}_n \otimes \hat{b}_n \otimes \hat{b}_n \otimes \hat{b}_n)_{ijkl}$, where $\alpha_\omega$ and $\alpha_{2\omega}$ are the first-order (linear) and second-order (hyper) polarizabilities, $\hat{b}_n$ is a unit vector that points along the *n*th bond, and all bond charges are assumed equal. Using this expression, we investigated how distortions along each of the 4 totally symmetric normal mode coordinates allowed in the $\bar{3}$ point group (i.e. the four basis functions $A_g^1$, $A_g^2$, $A_g^3$ and $A_g^4$ of its totally symmetric irreducible representations) change the individual $\chi_{ijkl}$ elements.

For simplicity, we considered only the nearest neighbor intralayer Cr-Te bonds and the nearest neighbor interlayer Cr-Cr bonds, which is reasonable because the states accessed by our photon energy ($2\hbar\omega = 3$ eV) are predominantly composed of Cr and Te orbitals[24,25]. Remarkably, our hyper-polarizable bond model shows that under a small distortion along the $A_g^1$ normal coordinate δ, which we implement by changing the $\hat{b}_n$ while keeping the $\alpha_\omega \alpha_{2\omega}$ values constant, only the *xxxz* and *yyyz* elements are affected (Fig. 4a), in perfect agreement with our observations below $T_{2D}$ (Fig. 3a). Since motion along $A_g^1$ deforms the Te octahedra and can bring the Cr-Te-Cr bond angle closer to 90° to strengthen $J_{ab}$, it is natural to associate this distortion with the development of FM in-plane spin correlations. This is further supported by neutron scattering experiments[7], which show a rise in magnetic diffuse scattering around $T_{2D}$ (inset Fig. 3a) indicative of a growing in-plane correlation length $\xi_{ab}$.

To uncover a mechanism that would exclusively affect the *zzzz* element below $T_{3D}$ (Fig. 3b), we note that the distortion along the $A_g^2$ normal coordinate involves a pure out-of-plane displacement of the Cr atoms. Although this motion does not change the $\hat{b}_n$ of the interlayer Cr-Cr bonds since they remain parallel to the *z*-axis, it will change their polarizabilities by virtue of their altered bond length. Assuming that it is these Cr-Cr bonds that primarily contribute to the observed changes at $T_{3D}$ (see Supplementary Note 6), our model indeed shows that tuning either $\alpha_\omega$ or $\alpha_{2\omega}$ of the Cr-Cr bond will exclusively affect the *zzzz* element (Fig. 4b). This naturally suggests an association of the $A_g^2$ distortion with the development and enhancement of FM interlayer spin correlations, and hence an identification of $T_{3D}$ as the 2D to 3D dimensional crossover temperature. Independent evidence for a structural distortion at $T_{3D}$ was also found via anomalies in the $E_g$ and $E_u$ phonons using impulsive stimulated Raman scattering (see Supplementary Note 7) and infrared absorption measurements[20] respectively (inset Fig. 3b), which likely arise from their nonlinear coupling to the $A_g^2$ distortion.

As a further consistency check, we note that one expects interlayer correlations to onset when $\xi_{ab}$ grows to a size where the total interlayer exchange energy becomes comparable to the temperature. In a mean field approximation, this condition is expressed as $T = N(T)J_c S(S+1)/3k_B$, where *N* is the number of in-plane correlated spins of magnitude *S* that are interacting with the next layer, $J_c$ is the interlayer Cr-Cr exchange and $k_B$ is Boltzmann's constant. Using the values of $\xi_{ab}(T)$ and $J_c$ determined from neutron scattering[7], we find a solution to the mean field equation at $T \sim 70$ K (see Supplementary Note 8), which is reasonably close to $T_{3D}$. Displacements along the remaining two $A_g^3$ and $A_g^4$ normal coordinates are found from our model to affect all 8 of the tensor elements (Fig. 4c,d) and are therefore not measurably induced at either $T_{2D}$ or $T_{3D}$. It is possible that they occur below $T_c$ where we observe all elements to change (Fig. 3), but details of LRO induced distortions are outside the scope of this work.

Our EQ SHG data and analysis taken together provide a comprehensive picture of how the quasi-2D Heisenberg ferromagnet CrSiTe$_3$ evades the Mermin-Wagner theorem via a multiple stage process to establish long-range spin order (Fig. 4e), and shows that interlayer interactions are vital to stabilizing LRO at such high temperatures. More generally, our results demonstrate that the nonlinear optical response is a highly effective probe of short-range spin physics and their associated totally symmetric magneto-elastic distortions, which are typically unresolvable by capacitance dilatometry[10] (Fig. 1a) or lower rank optical processes like linear reflectivity and Raman scattering due to their limited degrees of freedom (see Supplementary Note 9), and are challenging to detect by diffraction based techniques limited to pico-meter resolution[18]. This technique will be particularly useful for studying anisotropic or geometrically frustrated magnetic systems, which tend to display interesting short-range spin correlations. It will also be useful for uncovering magnetic ordering mechanisms in monolayer or few layer ferromagnetic and antiferromagnetic nanoscale flakes and devices[26–29], which are often unclear because of their inaccessibility by neutron diffraction. We anticipate that access to this type of information may offer new strategies to control magnetism based on manipulating SRO induced distortions through chemical synthesis, static perturbations or even out-of-equilibrium excitations[30].

**Methods**

Sample growth and characterization

The CrSiTe$_3$ crystals used in this study were grown using a Te self-flux technique [20]. High purity Cr (Alfa Aesar, 99.999%), Si (Alfa Aesar 99.999%) and Te (Alfa Aesar 99.999%) were weighed in a molar ratio of 1:2:6 (Cr:Si:Te) and loaded into an alumina crucible sealed inside a quartz tube. The quartz ampoule was evacuated and backfilled with argon before sealing. Plate-like crystals up to 5 mm thick with flat, highly reflective surfaces were then removed from the reaction crucible. X-ray diffraction (XRD) data collected on crushed crystals using an Emperyan diffractometer (Panalytical) confirmed the correct $R\bar{3}$, space group 148, $CrSiTe_3$ phase. Measurements of the temperature and field-dependence of the magnetization were carried out using a Magnetic Property Measurement System (MPMS, Quantum Design). The samples were mounted with the field applied parallel to the *ab*-plane of the crystals for magnetization and susceptibility measurements. The thermal expansion coefficient was measured using the Quantum Design dilatometer option in a PPMS DynaCool. Dilation was measured along the *c*-axis; sample thickness in this direction was 0.41 mm. Data were collected under a ramp rate of 0.1 K/min.

RA SHG measurements

Incident light with < 100 fs pulse width and 800 nm center wavelength was derived from a ti:sapph amplified laser system (Coherent RegA) operating at 100 kHz. Specular reflected second-harmonic light at 400 nm was selected using short-pass and narrow bandpass filters and measured with a two-dimensional EM-CCD camera (Andor iXon Ultra 897). Both the sample and detector remained fixed while the scattering plane is rapidly mechanically spun about the central beam axis. The angle of incidence was fixed at 10°. A detailed description of the RA SHG apparatus used can be found in Ref. 17. The fluence of the beam was maintained at ~340 µJ $cm^{-2}$ with a spot size of ~30 µm FWHM. The close agreement between the $T_c$ values measured using RA SHG and magnetic susceptibility indicates negligible average heating by the laser beam. Each complete RA pattern was acquired with a 5 min exposure time. Samples (~1 mm × 2 mm × 0.1 mm) were cleaved prior to measurement and immediately pumped down in an optical cryostat to a pressure better than $10^{-6}$ Torr.

# References

1. Gong, C. *et al.* Discovery of intrinsic ferromagnetism in two-dimensional van der Waals crystals. *Nature* **546**, 265–269 (2017).

2. Huang, B. *et al.* Layer-dependent ferromagnetism in a van der Waals crystal down to the monolayer limit. *Nature* **546**, 270–273 (2017).

3. Carteaux, V., Moussa, F. & Spiesser, M. 2D Ising-Like Ferromagnetic Behaviour for the Lamellar $Cr_2Si_2Te_6$ Compound: A Neutron Scattering Investigation. *EPL Europhys. Lett.* **29**, 251 (1995).

4. Li, X. & Yang, J. $CrXTe_3$ (X = Si, Ge) nanosheets: two dimensional intrinsic ferromagnetic semiconductors. *J. Mater. Chem. C* **2**, 7071–7076 (2014).

5. Chen, X., Qi, J. & Shi, D. Strain-engineering of magnetic coupling in two-dimensional magnetic semiconductor $CrSiTe_3$: Competition of direct exchange interaction and superexchange interaction. *Phys. Lett. A* **379**, 60–63 (2015).

6. Lin, M.-W. *et al.* Ultrathin nanosheets of $CrSiTe_3$: a semiconducting two-dimensional ferromagnetic material. *J. Mater. Chem. C* **4**, 315–322 (2015).

7. Williams, T. J. *et al.* Magnetic correlations in the quasi-two-dimensional semiconducting ferromagnet $CrSiTe_3$. *Phys. Rev. B* **92**, 144404 (2015).

8. Liu, B. *et al.* Critical behavior of the quasi-two-dimensional semiconducting ferromagnet $CrSiTe_3$. *Sci. Rep.* **6**, 33873 (2016).

9. Mermin, N. D. & Wagner, H. Absence of Ferromagnetism or Antiferromagnetism in One- or Two-Dimensional Isotropic Heisenberg Models. *Phys. Rev. Lett.* **17**, 1133–1136 (1966).

10. Zapf, V. S. *et al.* Direct measurement of spin correlations using magnetostriction. *Phys. Rev. B* **77**, 020404 (2008).

11. Shen, Y. R. Surface Second Harmonic Generation: A New Technique for Surface Studies. *Annu. Rev. Mater. Sci.* **16**, 69–86 (1986).

12. Fiebig, M., Pavlov, V. V. & Pisarev, R. V. Second-harmonic generation as a tool for studying electronic and magnetic structures of crystals: review. *JOSA B* **22**, 96–118 (2005).

13. Denev, S. A., Lummen, T. T. A., Barnes, E., Kumar, A. & Gopalan, V. Probing Ferroelectrics Using Optical Second Harmonic Generation. *J. Am. Ceram. Soc.* **94**, 2699–2727 (2011).

14. Zhao, L. *et al.* Evidence of an odd-parity hidden order in a spin–orbit coupled correlated iridate. *Nat. Phys.* **12**, 32–36 (2016).

15. Harter, J. W., Zhao, Z. Y., Yan, J.-Q., Mandrus, D. G. & Hsieh, D. A parity-breaking electronic nematic phase transition in the spin-orbit coupled metal $Cd_2Re_2O_7$. *Science* **356**, 295–299 (2017).

16. Matsubara, M. *et al.* Optical second- and third-harmonic generation on the ferromagnetic semiconductor europium oxide. *J. Appl. Phys.* **109**, 07C309 (2011).

17. Harter, J. W., Niu, L., Woss, A. J. & Hsieh, D. High-speed measurement of rotational anisotropy nonlinear optical harmonic generation using position-sensitive detection. *Opt. Lett.* **40**, 4671–4674 (2015).

18. Carteaux, V., Ouvrard, G., Grenier, J. C. & Laligant, Y. Magnetic structure of the new layered ferromagnetic chromium hexatellurosilicate $Cr_2Si_2Te_6$. *J. Magn. Magn. Mater.* **94**, 127–133 (1991).

19. Birss, R. R. *Symmetry and magnetism*. (North-Holland Pub. Co., 1964).

20. Casto, L. D. *et al.* Strong spin-lattice coupling in $CrSiTe_3$. *APL Mater.* **3**, 041515 (2015).

21. Wu, S. *et al.* The direct observation of ferromagnetic domain of single crystal CrSiTe$_3$. *AIP Adv.* **8**, 055016 (2018).

22. Powell, G. D., Wang, J.-F. & Aspnes, D. E. Simplified bond-hyperpolarizability model of second harmonic generation. *Phys. Rev. B* **65**, 205320 (2002).

23. Bauer, K.-D. & Hingerl, K. Bulk quadrupole contribution to second harmonic generation from classical oscillator model in silicon. *Opt. Express* **25**, 26567–26580 (2017).

24. Siberchicot, B., Jobic, S., Carteaux, V., Gressier, P. & Ouvrard, G. Band Structure Calculations of Ferromagnetic Chromium Tellurides CrSiTe$_3$ and CrGeTe$_3$. *J. Phys. Chem.* **100**, 5863–5867 (1996).

25. Sivadas, N., Daniels, M. W., Swendsen, R. H., Okamoto, S. & Xiao, D. Magnetic ground state of semiconducting transition-metal trichalcogenide monolayers. *Phys. Rev. B* **91**, 235425 (2015).

26. Li, X., Cao, T., Niu, Q., Shi, J. & Feng, J. Coupling the valley degree of freedom to antiferromagnetic order. *Proc. Natl. Acad. Sci.* **110**, 3738–3742 (2013).

27. Sivadas, N., Okamoto, S. & Xiao, D. Gate-Controllable Magneto-optic Kerr Effect in Layered Collinear Antiferromagnets. *Phys. Rev. Lett.* **117**, 267203 (2016).

28. Jiang, S., Li, L., Wang, Z., Mak, K. F. & Shan, J. Controlling magnetism in 2D CrI$_3$ by electrostatic doping. *Nat. Nanotechnol.* **13**, 549–553 (2018).

29. Huang, B. *et al.* Electrical control of 2D magnetism in bilayer CrI$_3$. *Nat. Nanotechnol.* **13**, 544–548 (2018).

30. Basov, D. N., Averitt, R. D. & Hsieh, D. Towards properties on demand in quantum materials. *Nat. Mater.* **16**, 1077–1088 (2017).

**Acknowledgements**

This work was supported by ARO MURI Grant No. W911NF-16-1-0361. D.H. also acknowledges support for instrumentation from the David and Lucile Packard Foundation and from the Institute for Quantum Information and Matter, an NSF Physics Frontiers Center (PHY-1733907). A.R. acknowledges support from the Caltech Prize Fellowship. The MRL Shared Experimental Facilities are supported by the MRSEC Program of the NSF under Award No. DMR 1720256; a member of the NSF-funded Materials Research Facilities Network. S.D.W authors acknowledge support from the Nanostructures Cleanroom Facility at the California NanoSystems Institute (CNSI). We thank Tom Hogan for performing the dilatometry measurements and Liangbo Liang, David Mandrus, Jan Musfeldt, Kai Xiao and Houlong Zhuang for helpful discussions.

**Author contributions**

A.R. and D.H. conceived the experiment. A.R. performed the optical measurements. A.R., D.H. and L.B. analysed the data. L.B. performed the classical Heisenberg model calculations. E.Z. and S.D.W. prepared and characterized the sample. A.R. and D.H. wrote the manuscript.

**Competing interests**

The authors declare no competing interests.

**Data availability**

The datasets generated are/or analyzed during the current study are available from the corresponding author on reasonable request.

**Additional information**

Supplementary information accompanies this paper. Correspondence and requests for materials should be addressed to D.H. (dhsieh@caltech.edu).

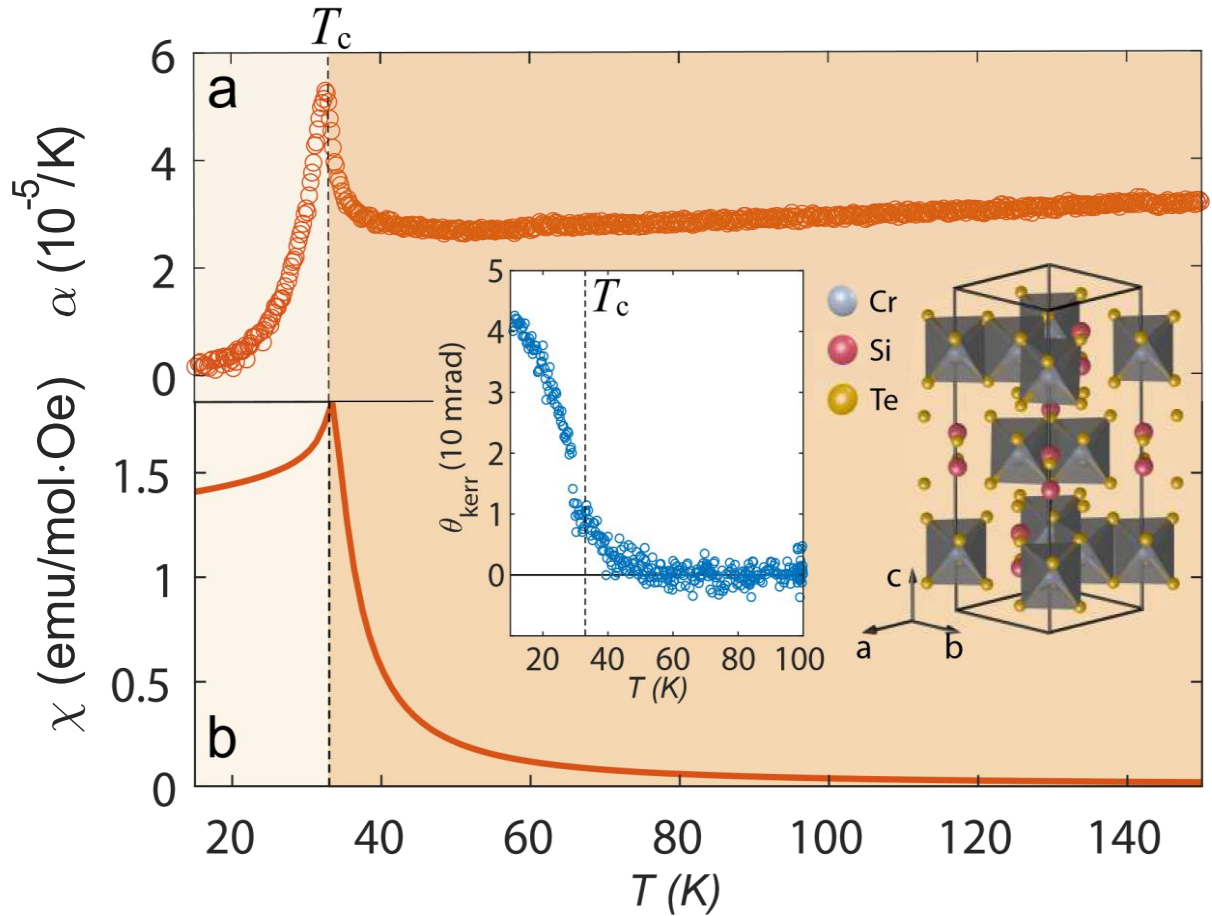

**Figure 1. Characterization of long-range spin ordering in CrSiTe$_3$.** Temperature dependence of the a, c-axis thermal expansion coefficient, b, magnetic susceptibility and (left inset) magneto-optical Kerr rotation angle from a bulk single crystal of CrSiTe$_3$ showing the clear onset of long-range magnetic order at $T_c$ ~ 31 K. Note the absence of any features above $T_c$ in all of these measurements. Right inset shows the crystal structure of CrSiTe$_3$. In the FM phase, the Cr$^{3+}$ spins all point parallel to the c-axis.

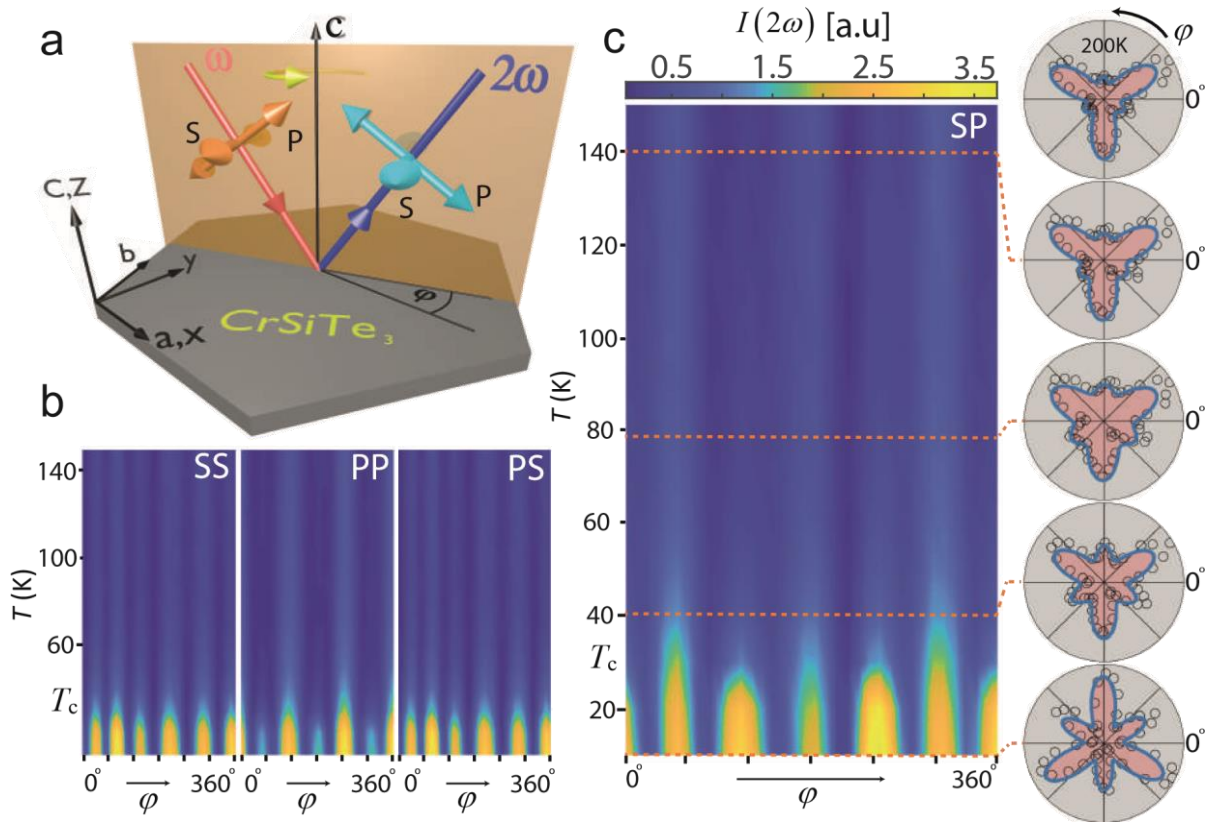

**Figure 2. Temperature dependence of the SHG rotational anisotropy. a,** Experimental geometry of the RA SHG experiment. The φ = 0° direction coincides with the crystallographic *a*-axis. **b,** Image plots of the raw temperature dependent RA SHG data acquired in SS, PP and PS polarization configurations, where the first and second labels denote the selected incident (ω) and detected (2ω) polarization components. The horizontal black line indicates the $T_c$ of our sample (see Fig. 1). **c,** A zoom-in on the RA SHG image plot acquired in SP configuration. Cuts at select temperatures displayed as polar plots (open circles) are shown to emphasize the changes in shape of the RA patterns taking place above $T_c$. Blue curves are best fits to the expected EQ SHG response from a $\bar{3}$ point group as described in the text.

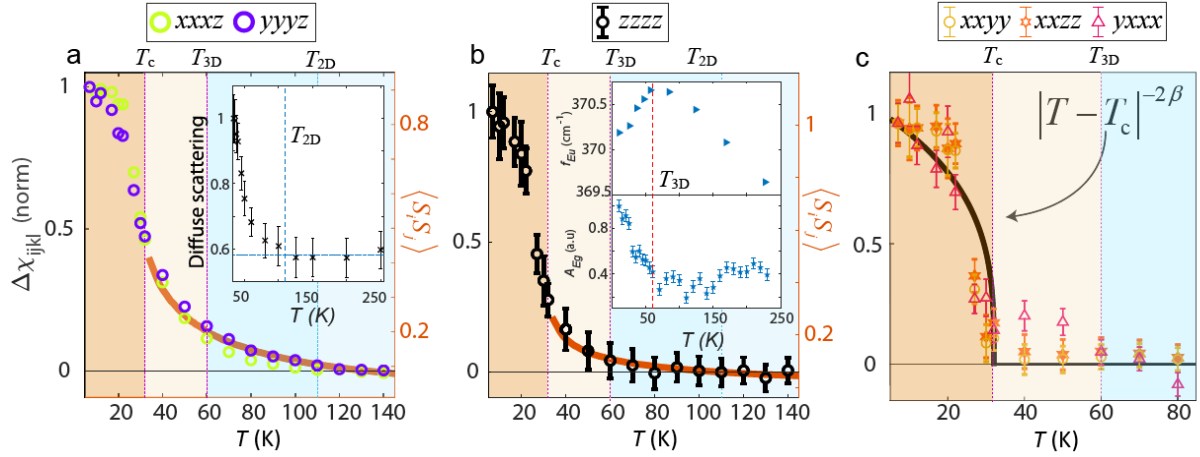

**Figure 3. Temperature evolution of individual susceptibility tensor elements.** The values of the **a,** *xxxz*, *yyyz*, **b,** *zzzz* and **c,** remaining elements of the SHG susceptibility tensor $\chi_{ijkl}$ at each temperature extracted through simultaneous fitting to all four polarization configurations of the RA data (see text). For every element, a high temperature background derived from the data above $T$ = 150 K was subtracted off. The background subtracted curves ($\Delta\chi_{ijkl}$) were then each normalized to their values at $T$ = 7 K. Error bars are the least squared errors of the fits, which are smaller than the data symbols in panel **a**. Solid lines in panels **a** and **b** are the calculated intralayer and interlayer spin correlator respectively. The line in panel **c** is the function $|T - T_c|^{-2\beta}$, where $\beta$ is the critical exponent of the magnetization. The two elements *zzxx* and *zzxy* are not displayed in **c** for clarity because their fitted values are very small, resulting in much larger error bars. However their trend follows that of the *xxyy*, *xxzz* and *yxxx* elements shown in **c** (see Supplementary Note 4). Diffuse magnetic neutron scattering data from $CrSiTe_3$ reproduced from Ref. [7] is shown in the inset of **a**. The inset of **b** shows the frequency of the $E_u$ phonon (top) and amplitude of the $E_g$ phonon (bottom), which are respectively reproduced from Ref. [20] and measured using impulsive stimulated Raman scattering (see Supplementary Note 7).

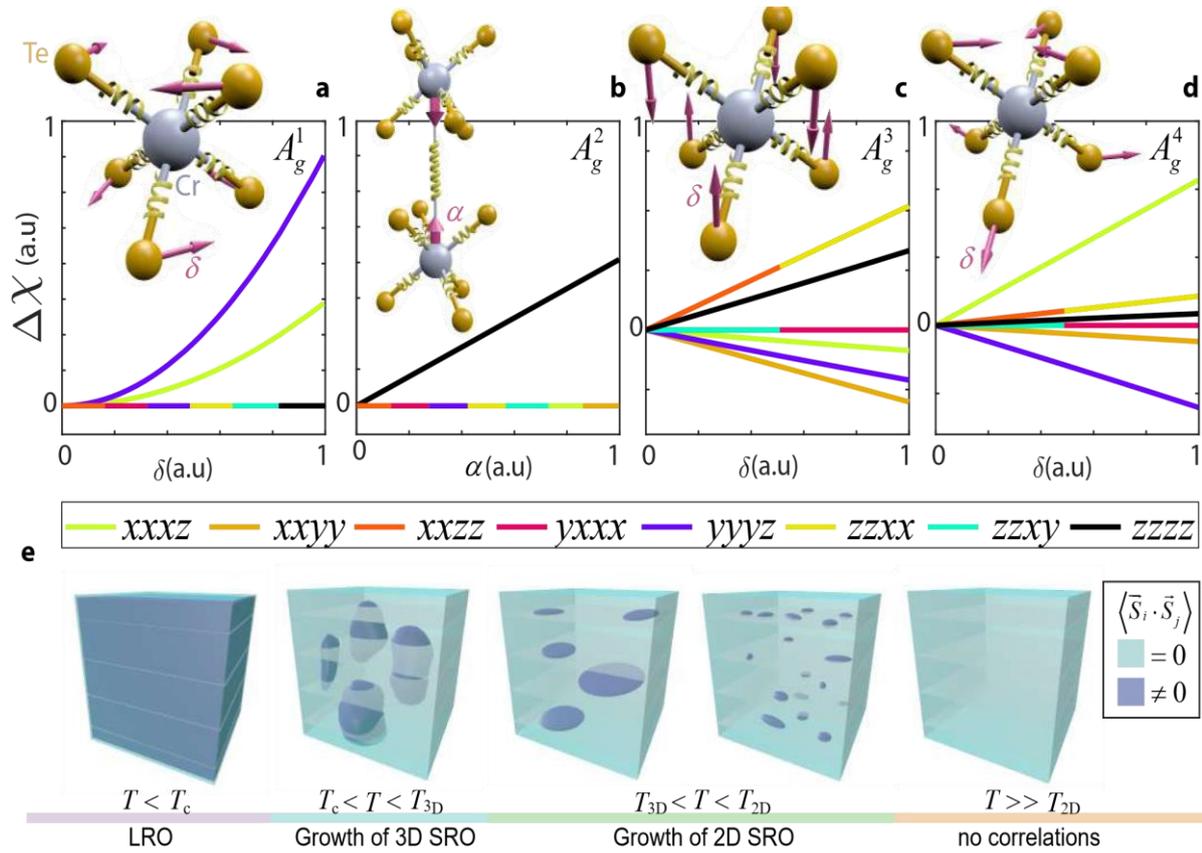

**Figure 4. Hyper-polarizable bond model and dimensional crossover.** Plots show hyper-polarizable bond model calculations of the change in each of the 8 allowed SHG susceptibility tensor elements relative to their undistorted values ($\Delta \chi_{ijkl}$) under totally symmetric distortions along the a, $A_g^1$, b, $A_g^2$, c, $A_g^3$ and d, $A_g^4$ normal coordinates. Schematics in the insets show the bonds (springs) used in our model and depict how each distortion is parameterized. Both vertical and horizontal axes are plotted on a linear scale. Tensor elements that change in the same way are represented as multi-colored curves. e, Illustration of the spin ordering process in CrSiTe$_3$ deduced from our data. The light blue regions represent an absence of spin-spin correlations and the dark blue regions represent FM correlations, which are dynamic above $T_c$ and static below $T_c$.

Supplementary Information for

# Dimensional crossover in a layered ferromagnet detected by spin correlation driven distortions

A. Ron et al.

Contents:

**Supplementary Note 1. Mathematical expressions for EQ induced RA-SHG patterns**

Expressions for the bulk EQ derived RA-SHG intensity $I$ from a crystal with $\bar{3}$ point group symmetry for the four experimental polarization configurations are given below, where $\theta = 10°$ is the experimental angle of incidence. The RA-SHG data were simultaneously fit to all four expressions to extract the values of the susceptibility tensor elements.

$$I_{PP} = \cos^2\theta\Big(\big(\sin^2\theta\chi_{zzxx}\cos^2\theta - \chi_{xxxz}\cos\theta\sin\theta\cos(3\varphi) - 2\chi_{zzxx}\sin^2\theta + \chi_{zzzz}\sin^2\theta - \tag{1}$$
$$-\frac{1}{2}\chi_{yyyz}\sin(2\theta)\sin(3\varphi)\big)^2 + \big(\chi_{xxxz}\cos\theta\cos(3\varphi)(\cos^2\theta - 2\sin^2\theta) -$$
$$-\sin\theta(3\chi_{xxyy}\cos^2\theta - 2\chi_{xxzz}\cos^2\theta + \chi_{xxzz}\sin^2\theta) +$$
$$+\chi_{yyyz}(\cos^3\theta - \sin\theta\sin(2\theta))\sin(3\varphi)\big)^2\Big)$$

$$I_{PS} = \big(\chi_{yyyz}\cos\theta\cos(3\varphi)(\cos\theta - 2\sin^2\theta) + \sin\theta(\chi_{yxxx}\cos^2\theta + 2\chi_{zzxy}\cos^2\theta - \chi_{zzxy}\sin^2\theta) + \tag{2}$$
$$+\chi_{xxxz}(-\cos^3\theta + \sin\theta\sin(2\theta))\sin(3\varphi)\big)^2$$

$$I_{SP} = \cos^2\theta\big(\chi_{xxxz}\cos\theta\cos(2\varphi) + \chi_{xxyy}\sin\theta + \chi_{yyyz}\cos\theta\sin(3\varphi)\big)^2 + \tag{3}$$
$$+\sin^2\theta\big(\chi_{zzxx}\cos\theta + \chi_{xxxz}\cos(3\varphi)\sin\theta + \chi_{yyyz}\sin\theta\sin(3\varphi)\big)^2$$

$$I_{SS} = \big(\chi_{yxxx}\sin\theta + \cos\theta(\chi_{yyyz}\cos(3\varphi) - \chi_{xxxz}\sin(3\varphi))\big)^2 \tag{4}$$

## Supplementary Note 2. Best fits to alternative SHG processes

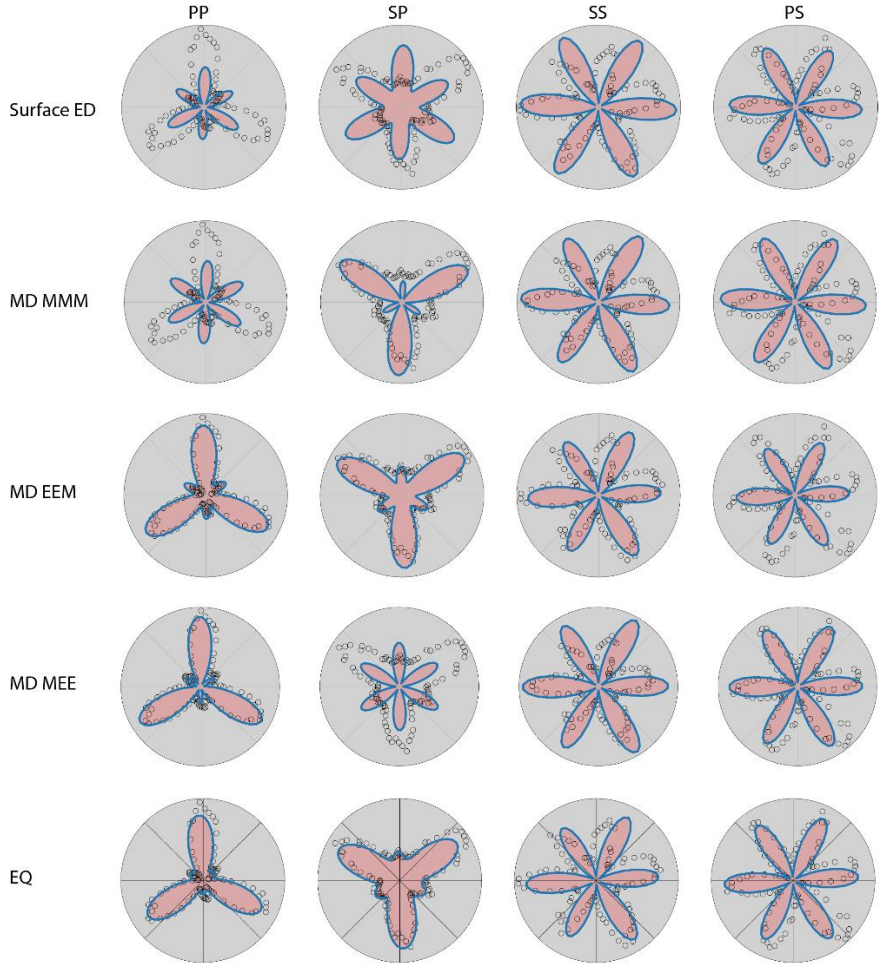

Supplementary Figure 1. Best fits of the RA-SHG data (circles) at $T = 200$ K to the following SHG processes (blue curves): (from top to bottom) surface ED, bulk MD (MMM), bulk MD (EEM), bulk MD (MEE) and bulk EQ. Simultaneous fits were made to the data in four different polarization configurations: PP, SP, SS and PS. The intensity in the SS and PS channels is ~ 50 % smaller than in the PP and SP channels, which is the reason for the larger scatter in those data sets.

We examined all of the leading order contributions to SHG allowed in a centrosymmetric crystal: the surface electric-dipole (ED) process given by $P_i = \chi_{ijk}^{EEE} E_j E_k$, three bulk magnetic-dipole (MD) processes given by $M_i = \chi_{ijk}^{MMM} H_j H_k$, $P_i = \chi_{ijk}^{EEM} E_j H_k$ and $M_i = \chi_{ijk}^{MEE} E_j E_k$, and a bulk electric-quadrupole (EQ) process given by $P_i = \chi_{ijkl}^{QEE} E_j \nabla_k E_l$, where $P$ and $M$ are the induced polarization and magnetization, and $E$ and $H$ are the incident electric and magnetic fields respectively. Representative best fits of each of these processes to our RA-SHG data are shown in Supplementary Figure 1. Clear discrepancies are observed for the surface ED and bulk MMM and MEE processes and are therefore immediately ruled out. Discrepancies in the bulk EEM fits also exist but are more subtle. Most notably, the small lobes in both PP and SP polarizations are not adequately captured over most of the temperature range of the experiment. Although these discrepancies in the bulk EEM fits are difficult to notice in the polar plots shown in Supplementary Figure 1, they are clear when plotted in Cartesian form (Supplementary Figure 2).

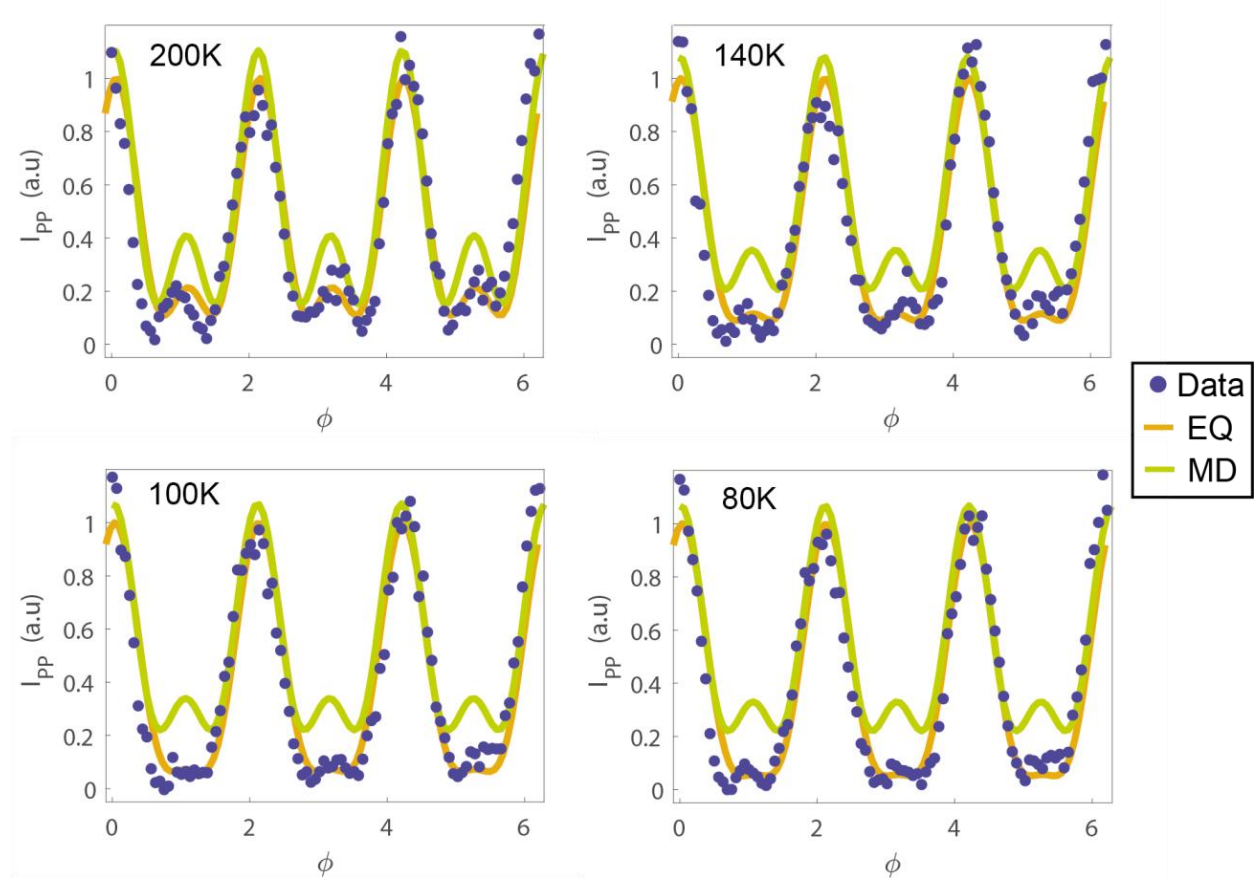

Supplementary Figure 2. Best fits to the RA-SHG data (blue circles) for the EQ (orange line) and MD EEM (green line) processes at various temperatures for the PP configuration. It is clear that the EQ process provides good fits to the data whereas the MD process does not.

**Supplementary Note 3. High temperature background subtraction procedure and fits**

To obtain the curves shown in Figure 3 of the main text, we first extracted the temperature dependence of each susceptibility tensor element by fitting the expressions given in Supplementary Note 1 to the raw RA-SHG data. For each tensor element independently, we then subtracted a weakly linear background using the data above 150 K where the shapes of the RA-SHG patterns have ceased changing (i.e. far above where magnetic correlations start to affect our data). In Supplementary Figure 3 we explicitly show that over the temperature range used for estimating the background, there is indeed no measureable change in shape of the RA-SHG patterns. The direct fit results before any background subtraction or normalization is performed are shown in Supplementary Figure 4.

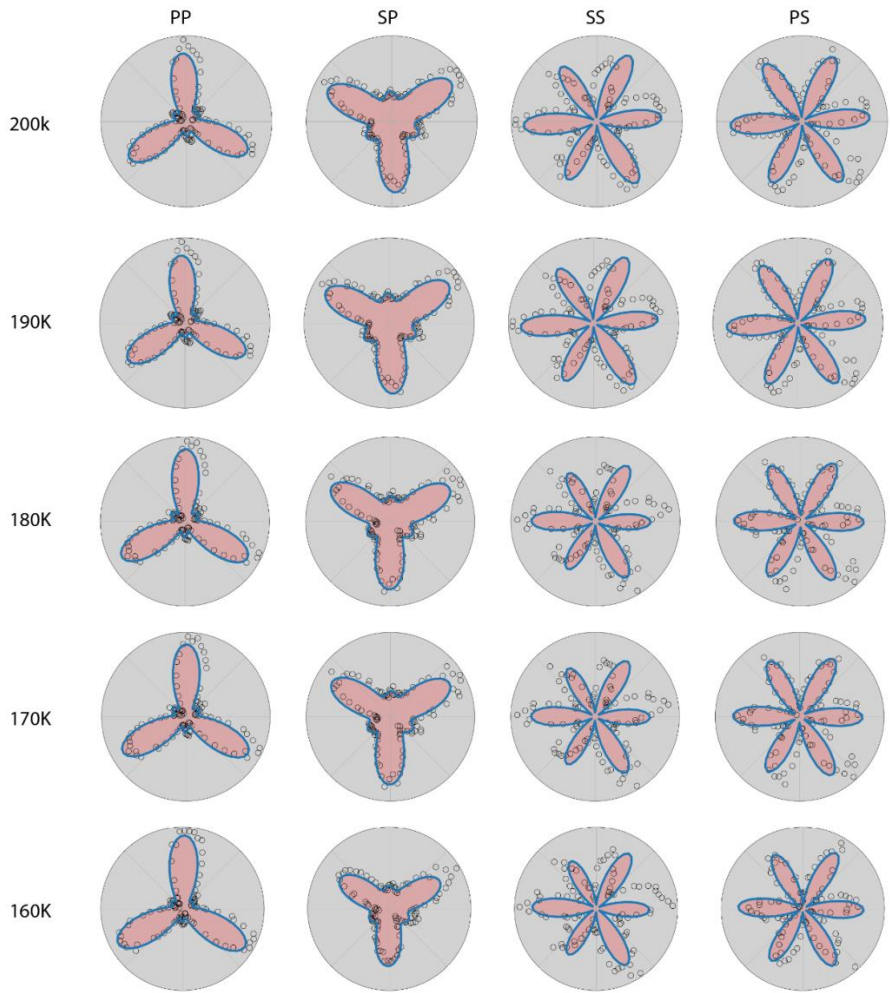

Supplementary Figure 3. RA-SHG data (circles) and EQ fits (blue curves) at various temperatures above 150 K.

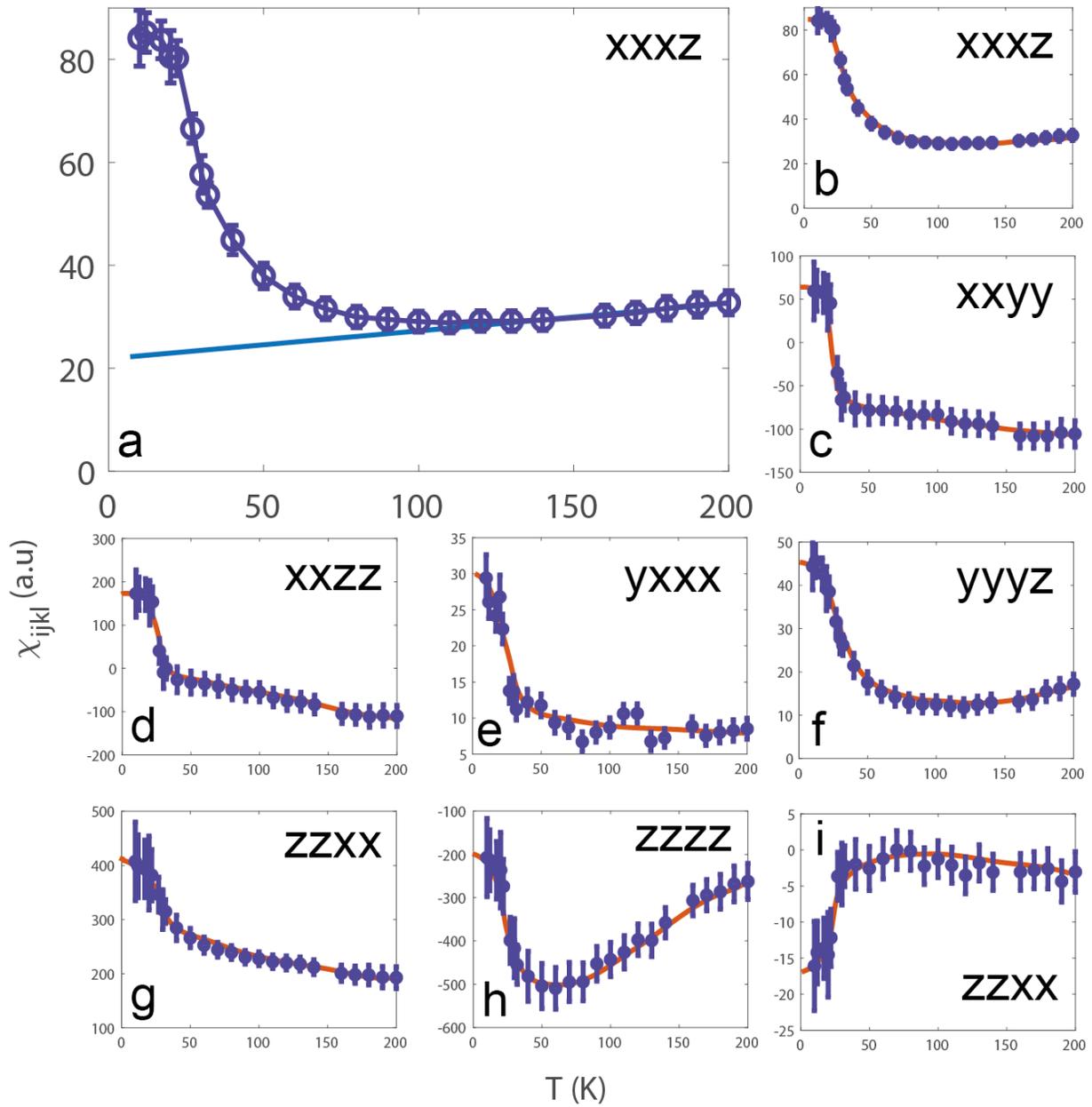

Supplementary Figure 4. Temperature dependence of the various tensor elements obtained directly from fitting the RA-SHG data. Error bars are the least squared errors of the fits. The blue curve in **a** is a linear fit to the data above 150 K, which is then subtracted off to produce the $\Delta\chi_{ijkl}$ curves shown in Fig. 3 of the main text. The same procedure was independently performed for all 8 tensor elements. The results of the fitting procedure without any background subtraction or normalization are shown in panels **b-i** for all tensor elements. Red lines are guides to the eye.

## Supplementary Note 4. Temperature dependence of $\chi_{zzxx}$ and $\chi_{zzxy}$

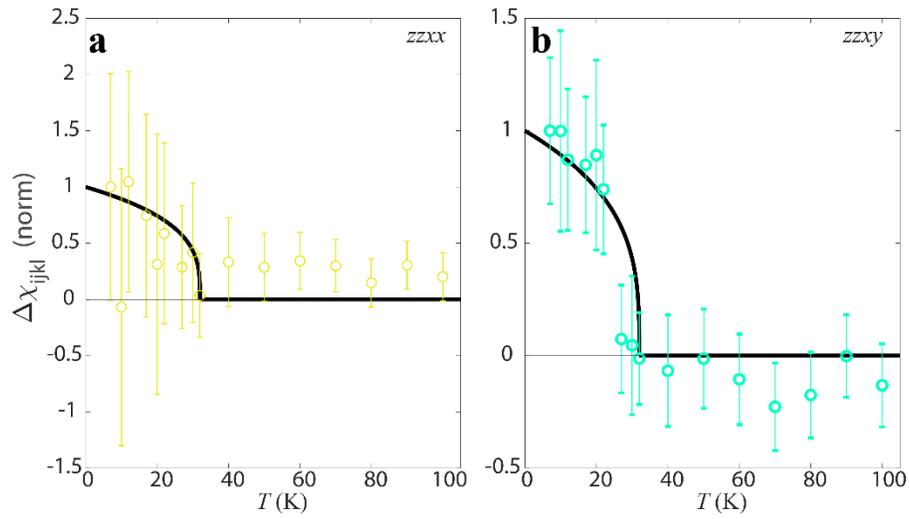

Supplementary Figure 5. Normalized and background subtracted temperature dependence of the **a,** *zzxx* and **b,** *zzxy* tensor elements, which are not shown in Fig. 3c of the main text because their absolute values are small, leading to much larger error bars than the other tensor elements. These two elements behave like the *xxyy*, *xxzz* and *yxxx* elements, turning up only below $T_c$. The black lines are proportional to the function $|T - T_c|^{-2\beta}$ where $\beta$ is the critical exponent of the magnetization.

**Supplementary Note 5. Classical Heisenberg model calculations**

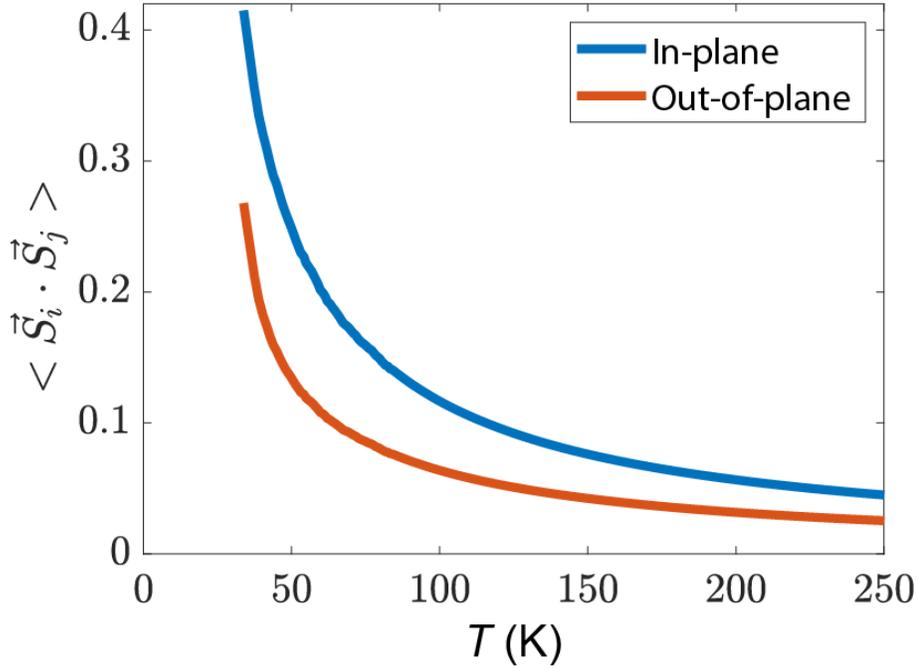

Supplementary Figure 6. Temperature dependence of the nearest neighbor intralayer (blue) and interlayer (orange) spin correlators calculated by solving the classical Heisenberg model in a large-$N$ approximation using the experimental parameters for CrSiTe$_3$.

To understand the temperature dependence of the short-range spin correlations in CrSiTe$_3$, we calculated the nearest neighbor correlations from a classical Heisenberg model, treating the spins as vectors of length $S = 3/2$ and using the exchange parameters from Ref. [7]. To calculate the correlations, we used a large-$N$ approximation, expected to be reasonably accurate above the critical temperature. Specifically, the fixed length constraint is implemented on average via a Lagrange multiplier $\lambda(T)$, which is self-consistently determined at each temperature. The spin correlations were then obtained as the Fourier transform of the inverse exchange interaction matrix, shifted appropriately by $\lambda(T)$. This approach gives $T_c = 33.8$ K, which agrees excellently with experiment and serves as a further self-consistency check of the dimensional crossover picture. Supplementary Figure 6 shows the calculated temperature dependence of the nearest neighbor intralayer and nearest neighbor interlayer spin correlator, which shows a sub-linear temperature dependence that is consistent with that of the *xxxz*, *yyyz* and *zzzz* susceptibility tensor elements shown in Figure 3 of the main text.

# Supplementary Note 6. Effect of $A_g^2$ distortion on Cr-Te bond contribution to $\Delta\chi_{ijkl}$

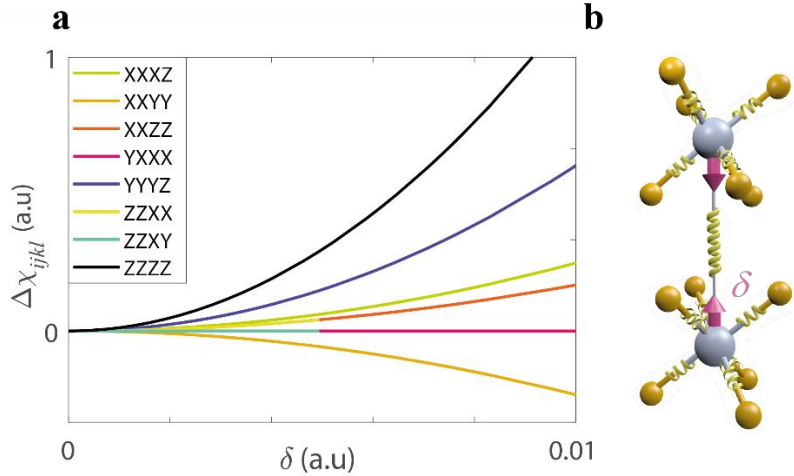

Supplementary Figure 7. **a,** Change in the susceptibility tensor elements due to a displacement along the $A_g^2$ normal coordinate calculated using the hyper-polarizable bond mode including only the Cr-Te bonds. **b,** Schematic of the displacement δ of the Cr atoms.

In Figure 4b of the main text, we showed how a displacement along the $A_g^2$ normal coordinate affects the interlayer Cr-Cr bond contribution to $\chi_{ijkl}$. Here we show how this distortion affects the intralayer Cr-Te bond contribution to $\chi_{ijkl}$ through a modification of the $\hat{b}_n$ associated with the Cr-Te bonds. As shown in Supplementary Figure 7, it is again the *zzzz* element that is primarily affected, consistent with our experiments (Fig. 3b). However, in this case the *xxxz, yyyz, xxzz, zzxx* and *xxyy* elements are also affected to a smaller degree, which we cannot resolve experimentally. Therefore we conclude that the Cr-Cr bond contribution dominates over the Cr-Te contribution for the $A_g^2$ distortion.

**Supplementary Note 7. Details of impulsive stimulated Raman scattering measurements**

To independently verify the presence of a structural distortion at $T_{3D}$, we performed impulsive stimulated Raman scattering (ISRS) measurements on CrSiTe$_3$ to track the behavior of Raman active phonons as a function of temperature. In our experiment, the sample was excited using an ultrashort (< 100 fs) optical pump pulse at a wavelength of 1200 nm, which excites coherent phonons through the ISRS mechanism. Phonon vibrations were then resolved in the time-domain by measuring the instantaneous polarization rotation angle θ of a reflected optical probe pulse at a wavelength of 800 nm as a function of time-delay $t$. As shown in Supplementary Figure 8, traces of θ($t$) exhibit clear oscillations due to coherent phonons and are dominated by two frequency components at ~3.7 THz and ~2.8 THz, which correspond to the frequencies of the $A_g$ and $E_g$ modes respectively. The amplitudes of the modes were determined by the peak intensities of the Fourier transform (inset Supplementary Figure 8) and tracked as function of temperature. A clear anomaly in the amplitude of the $E_g$ mode was observed at $T_{3D}$ (inset Fig. 3b). No clear anomaly at $T_{3D}$ was found in the $A_g$ mode, likely indicating a weaker nonlinear coupling to the $A_g^2$ distortion.

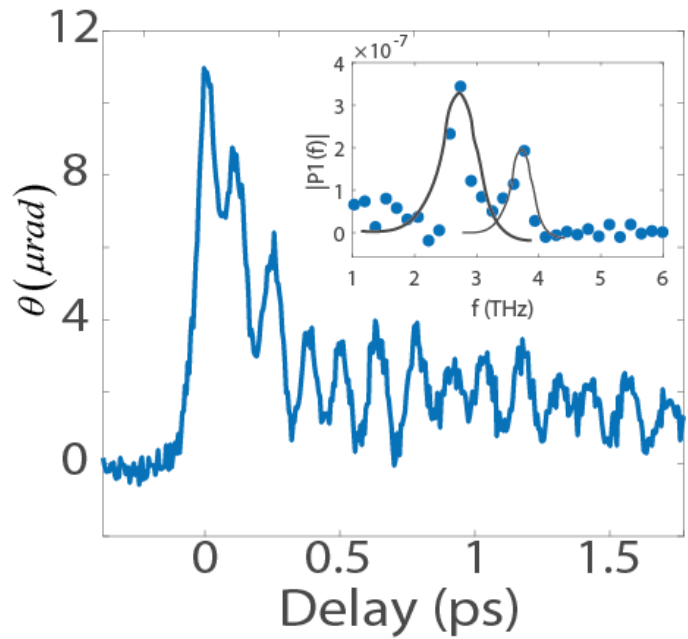

Supplementary Figure 8. Time-resolved polarization rotation of CrSiTe$_3$ acquired at $T$ = 15 K showing a beat pattern coming from the $E_g$ and $A_g$ phonon modes. The two frequency components are more clearly resolved in the Fourier transform (inset).

## Supplementary Note 8. Mean-field estimate of $T_{3D}$

We expect the interlayer correlations to onset when the total interlayer exchange energy becomes comparable to the temperature. In a mean field approximation this condition is expressed as $T = N(T)J_c S(S+1)/3k_B$. where $N$ is the number of in-plane correlated spins of magnitude $S$ that are interacting with the next layer, $J_c$ is the interlayer Cr-Cr exchange and $k_B$ is Boltzmann's constant. When counting $N(T)$, it is important to note that there are two inequivalent Cr sites due to the $ABC$ type stacking (Supplementary Figure 9 inset). One type of site has a Cr ion directly above and a Si pair directly below, while the other type of site has the reversed situation. Each site occupies one of the two sub-lattices of the honeycomb structure. When calculating the total interlayer exchange energy one should therefore count only Cr ions on the same type of site (i.e. sub-lattice, see Supplementary Figure 9 inset) because only they couple to the same adjacent layer. The Cr ions that sit on a different sub-lattice are coupled through $J_c$ to a different layer. Since it is only the second nearest neighbors (2NN) that sit on the same type of site, when the in-plane spin correlation length $\xi_{ab}(T) < d_{2NN}$, $N(T) = 1$ (only the Cr ion itself), and when $d_{2NN} < \xi_{ab}(T) < d_{3NN}$, $N(T) = 7$ (distances defined in Supplementary Figure 9 inset). Using the values of $\xi_{ab}(T)$ and $J_c$ determined from neutron scattering [7], we find that $\xi_{ab}(T)$ reaches $d_{2NN}$ at $T \sim 65$ K (Supplementary Figure 9). Plugging $N(65\ K) = 7$ into the mean field equation gives a temperature of $\sim 75$ K, which is reasonably close to 65 K. Thus we estimate the mean field condition to be satisfied at $T_{3D} = 70$ K, in reasonable agreement with our observed value of $T_{3D} \sim 60$ K.

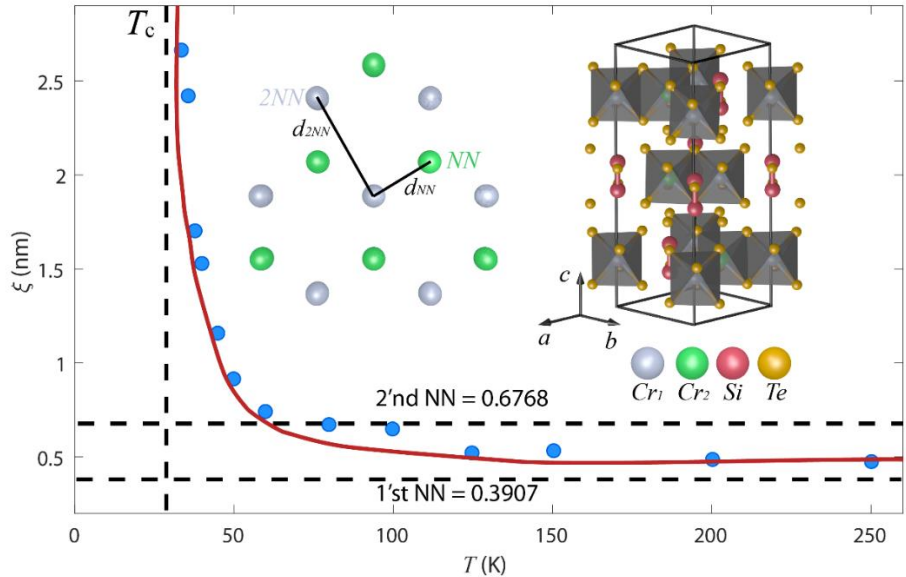

Supplementary Figure 9. 2D Coherence length as function of temperature [digitized from Ref. [7] Fig. 2(b)]. The blue dots are data points and the red line is a guide to the eye. Vertical dashed line marks $T_c$ and the horizontal dashed lines mark the 1'st and 2'nd nearest neighbors distances. Inset (right) is an illustration of the crystal structure highlighting the two inequivalent Cr sites in silver and green. Inset (left) is an illustration of the honeycomb structure of the Cr atoms in a single layer. The two inequivalent sites occupy the two honeycomb structure sub-lattices. Each Cr atom has 3 nearest neighbors (NN) and 6 next nearest neighbors (2NN).

**Supplementary Note 9. Bond model results for a rank-2 tensor response**

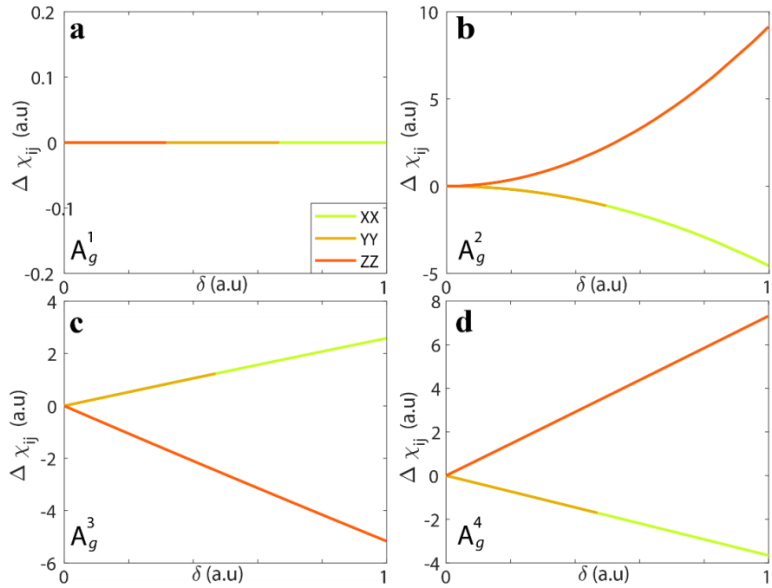

Supplementary Figure 10. Changes to a second rank polar tensor (reduced under point group $\bar{3}$) induced by totally symmetric distortions along the 4 normal coordinates described in the main text.

To understand how totally symmetric distortions along the 4 normal coordinates $A_g^1 \rightarrow A_g^4$ affect a rank-2 polar tensor $\chi_{ij}$ rather than a rank-4 polar tensor $\chi_{ijkl}$ as studied in the main text, we performed analogous hyper-polarizable bond model calculations for $\chi_{ij}$. For a crystal with $\bar{3}$ point group symmetry, there are 3 non-zero elements in $\chi_{ij}$, of which only 2 are independent (*xx* = *yy*, *zz*). The hyper-polarizable bond model calculations in Supplementary Figure 10 show that for the $A_g^1$ distortion, no elements change whereas for the other 3 distortions, all elements change. This shows that a rank-2 tensor response does not have enough degrees of freedom to differentiate between the distortions and one must appeal to a higher-rank nonlinear response.

We note that even though techniques such as Raman, IR or time-resolved optical spectroscopy are "mode specific", they are sensitive to phonon modes and not to static distortions like shown in our work. While it is reasonable to assume that a static distortion in some basis function will have an effect on the corresponding phonon (e.g. a change in its amplitude, frequency or lifetime), such effects commonly arise from non-linear couplings between the modes, or between the modes and other crystal degrees of freedom through for example spin-phonon or electron phonon coupling. This is in fact the reason for the changes at $T_{3D}$ shown in the inset of Figure 3b. Therefore the interpretation of such measurements (via frequency, amplitude or lifetime) cannot be associated directly to a static change in the corresponding basis function. A clear example can be taken from CrSiTe$_3$, where a change in the frequency of 3 IR modes, which lack a center of inversion, was reported in Ref. [20] to occur close to $T_c$, even though the crystal obviously remains centrosymmetric.